\documentclass[reprint,rmp,amsmath,amssymb,aps,floatfix]{revtex4-2}

\usepackage{graphicx}
\usepackage{dcolumn}
\usepackage{bm}
\usepackage{subcaption}
\usepackage[font={small,rm},labelfont=bf,textfont={it}]{caption}

\begin{document}
\preprint{APS/123-QED}

\title{Investigation into length scale dominance in critical black hole formation} 

\author{Cole Kelson-Packer}
\email{u0980902@utah.edu}
\author{John Belz}
\email{belz@physics.utah.edu}
\affiliation{
Department of Physics and Astronomy, University of Utah \\
Salt Lake City, Utah 84112, USA
}
\date{\today}

\begin{abstract}
The critical formation of low-mass black holes is a historical cornerstone of numerical general relativity, with important implications in cosmology for censorship conjectures and the production of primordial black holes (PBHs). Concurrent with the surge in black hole observational physics in recent years has been an increased interest in these subjects. Critical formation is often suggested as a mechanism for PBH production, but it is possible that the existence of different types of critical processes accompanying more realistic scenarios may affect this conclusion more than has been considered thus far. This paper numerically investigates, as a toy model, the interplay of multiple near-critical fields in the collapse of spherically symmetric scalar fields. It is found that a combination of type~I and type~II near-critical fields results in a kind of competition between their respective critical evolutions. A heuristic explanation for this phenomenon is proposed employing ideas from the theory of dynamical systems.
\end{abstract}

\maketitle

\section{Introduction}
\label{sec:intro}
\par Critical phenomena in black hole formation is one of the classic numerical results of General Relativity in the strongly interacting regime, dating back to Choptuik's~\cite{choporiginal} seminal paper on self-gravitating massless scalar fields. Similar critical phenomena without a mass gap have been discovered for a variety of different matter configurations, such as axially symmetric gravitational waves~\cite{abraev} and Yang-Mills fields~\cite{chopy}. Critical phenomena with a mass gap have also been discovered, for example in the study of massive scalar fields~\cite{othermassive}, and other ``hair,'' such as charge and angular momentum, exhibit critical behavior as well~\cite{gundele,hpele,Petrykele,gundang}. A larger collection of results may be found gathered in a review by Gundlach~\cite{gundlachrev}. Generally speaking, critical phenomena is a fine way of illustrating the richness of behavior accompanying the nonlinear nature of Einstein's equations.
\par Most studies of black hole critical phenomena, with recent exceptions~\cite{panic,kaindouble}, have considered a single type of constituent matter, and focus on initial data of that specific type belonging to various single-parameter families. This works well enough for illustrating criticality and quasiuniversality, as the mass (or whatever quantity is of interest) depends on a difference of the parameter, while universality is suggested by the similarity of behavior for a variety of parametrized initial data. Conclusions from such investigations have been considered sufficient for most applications of the theory: cosmic censorship conjectures are adequately probed by what are essentially toy models~\cite{c4, wildwald}, whereas mechanisms for producing primordial black holes are modeled upon the density fluctuations of dominating matter sources~\cite{malkuth,nj2,nj3}. 
\par What has been given less consideration is the implications that more realistic mixed matter configurations could have for criticality. Loosely speaking, critical phenomena in general relativity are the manifestation of the existence of different basins of attraction in the phase space of solutions to Einstein's equations, all associated with critical solutions of varying codimension~\cite{gundreview}. Different types of criticality, however, may be affiliated with different critical solutions: the evolution of massless scalar fields is influenced by the existence a self-similar spacetime~\cite{gundconst1,gundconst2}, whereas massive scalar fields exhibiting a critical mass gap are associated with metastable soliton stars~\cite{othermassive}. From previous numerical studies~\cite{othermassive} there would seem to exist interactions between the effects of different critical spacetimes: for critical massive scalar fields, a mass gap emerges and criticality shifts from type~II to type~I when the characteristic length scale of the initial field becomes sufficiently large. It is conceivable that the time evolution of more general composite configurations may be significantly affected by several critical solutions.
\par We test this idea in this paper. Considering spherically symmetric matter configurations that feature both a massless and a massive scalar field, we show using a two-parameter sample of initial conditions not only that three different phases of evolution behavior (corresponding to two collapse timescales and asymptotic dispersal) are displayed for this multifield content, but that competition between the influence of the critical solutions associated to the two individual fields affects the critical evolution of these spacetimes. Our results are similar in nature to recent, earlier work by Gundlach, Baumgarte, and Hilditch~\cite{panic}, resembling an alternative scenario they advance. An intriguing particularity highlighted by our results concerning the inhibiting effect of multicritical configurations, however, suggests that a more nuanced approach may be necessary when considering the application of critical phenomena to more realistic scenarios. 

\section{Methods}
\label{sec:methods}
\par Throughout we use Einstein summation convention and set $c$ and $8\pi G$ to unity for convenience.
\par We employ the polar-areal gauge for our metric. The toy model we use for illustrating our conceptual idea consists of a pair of spherically symmetric scalar fields minimally coupled to gravity. One field is massless, exhibiting type~II critical collapse when taken alone, as in Choptuik's original paper~\cite{choporiginal}. The other field is massive, with the mass, characteristic length scale, and initial conditions taken such that type~I critical collapse would be exhibited if it were evolved on its own~\cite{othermassive}.
\subsection{Matter evolution}
At the most general level, the Lagrangian for a collection of minimally coupled scalar fields is
\begin{equation}
L = \tfrac{1}{2}\bigtriangledown _{\mu}\Phi _i\bigtriangledown ^{\mu}\Phi _i-V(\Phi _i).
\end{equation}
\par With two fields, one massless, the other massive, and no other potential,
\begin{equation}
L = \tfrac{1}{2} \bigtriangledown _{\mu}\Phi _1\bigtriangledown ^{\mu}\Phi _1+\tfrac{1}{2}\bigtriangledown _{\mu}\Phi _2\bigtriangledown ^{\mu}\Phi _2-\tfrac{1}{2}m_1 ^2\Phi _1^2.
\end{equation}
\par A quick application of Euler-Lagrange yields the naive equations of motion:
\begin{eqnarray} \label{eq:fieldevolve}
\bigtriangledown _{\mu}\bigtriangledown ^{\mu}\Phi_1+m_1 ^2\Phi _1&&=0,\nonumber \\
\bigtriangledown _{\mu}\bigtriangledown ^{\mu}\Phi_2&&=0.
\end{eqnarray}
\par With our choice of metric, 
\begin{equation}
g_{\mu \nu} = diag(-\alpha ^2, a^2, r^2, r^2\sin ^2(\theta)),
\end{equation}
the Laplacian may be readily expanded:
\begin{equation}
\bigtriangledown _{\mu}\bigtriangledown ^{\mu}\Phi_j = \frac{1}{\alpha a}\partial _t(\frac{a}{\alpha}\partial _t\Phi _j)-\frac{1}{\alpha ar^2}\partial _r(\frac{\alpha r^2}{a}\partial _r\Phi _j).
\end{equation}
\par Defining the following auxiliary quantities,
\begin{equation}
\Pi _i\equiv \frac{a}{\alpha}\partial _t\Phi _i,\mbox{    }\Psi _i\equiv \partial _r\Phi _i,
\end{equation}
\par the equations (~\ref{eq:fieldevolve}) split into three pairs:
\begin{eqnarray}
\partial_t \Phi_i & = & \frac{\alpha}{a}\Pi_i, \hspace{4.2cm} i = 1,2, \nonumber \\
\nonumber \\
\partial_t \Psi_i & = & \partial_r \left( \frac{\alpha}{a} \Pi_i \right), \hspace{3.3cm} i = 1,2, \nonumber \\
\partial_t \Pi_i & = & \frac{1}{r^2} \partial_r \left( \frac{\alpha r^2}{a} \Psi_i \right) - \alpha a m^2_i \Phi_i, \hspace{0.85cm} i = 1,2, \nonumber \\
 & & \hspace{4.8cm} m_2 = 0. 
\end{eqnarray}
\par On the numerical level, the usual accommodations (see e.g. ~\cite{alucart}) are made for the third equation above so as to facilitate better behavior at the origin:
\begin{eqnarray}
\partial_t \Pi_i & = & 3\frac{\partial}{\partial r^3} \left( \frac{\alpha r^2}{a} \Psi_i \right) - \alpha a m^2_i \Phi_i \hspace{0.85cm} i = 1,2, \nonumber \\
 & & \hspace{4.8cm} m_2 = 0.
\end{eqnarray}
\par Simple radiating (Sommerfeld) boundary conditions are taken for the $\Psi _i$s and $\Pi _i$s, while the $\Phi _i$s are evolved using the same equation as above at the outer boundary. Specifically, we use, as a fair approximation,
\begin{eqnarray}
\partial _t \Pi_i & = &-\Pi_i/r-\partial _r\Pi_i, \nonumber \\
\Psi_i & = & -\Pi_i-\Phi_i/r. \nonumber
\end{eqnarray}
\par This is sufficient, but imperfect: for large t, apparent convergence may eventually degrade, even for dispersing initial conditions. 
\subsection{Metric evolution}
\par In the polar areal gauge the surface area of a sphere is held constant. This implies that the coefficient of the spherical area element $d\Omega $ is unity and that all components of intrinsic curvature $K_{ij}$ are zero except for the radial-radial component~\cite{alucart}. With the shift $\beta $ also chosen to be trivial, the ADM evolution equations simplify greatly. These choices mean that the only dynamical components of the metric are the lapse $\alpha $ and the radial-radial component $a$, which can be shown to satisfy the following equations:
\begin{equation} \label{eq:aevolve}
\partial _ra = \frac{a}{2}\left[ \frac{1-a^2}{r}+\frac{r}{2} \sum _{i=1}^{2} (\Pi _i^2+\Psi _i^2+m_i^2a^2\Phi _i^2)\right],
\end{equation}
\begin{equation} \label{eq:alevolve}
\partial _r\alpha = \alpha \left[ \frac{\partial _ra}{a}+\frac{a^2-1}{r}-\frac{m_1^2r}{2}a^2\Phi _1^2 \right].
\end{equation}
\par The first equation above arises from our demands upon the intrinsic curvature, while the second ultimately derives from the Hamiltonian constraint. The momentum constraint, meanwhile, yields the expression
\begin{equation}
0=M\equiv \alpha \tfrac{r}{2}(\Pi _1\Psi _1+\Pi _2\Psi _2)-\partial _ta,
\end{equation}
\noindent whose numerical deviation from exact satisfaction we use to monitor convergence. The time derivative in the last term of the above expression is evaluated using sixth-order centered finite difference, since fourth-order is expected.
\subsection{Numerical technique}
\par The basic underlying techniques we employ are standard, and may be found in most textbooks on numerical relativity, e.g.~\cite{alucart,baumpiro}. Starting from an initial configuration for the scalar fields belonging to a two-parameter space, we integrate equations (\ref{eq:aevolve}) and (\ref{eq:alevolve}) above using fourth-order Runge-Kutta to obtain $\alpha$ and $a$, demanding that $a=1$ at the origin. At the outer boundary we take $\alpha=1/a$, after obtaining $a$ via a basic fourth-order extrapolation. Having calculated the metric components, we are able to evolve the field components in time (also with fourth-order Runge-Kutta), allowing the reintegration $a$ and $\alpha$ at the next time step and subsequent repetition. Sixth-order dissipation is employed~\cite{GKO}, without which spurious oscillations develop coincident with the origin and the apparent horizon.
\par Our numerical grid extends radially to $r~=400$. This limit was chosen for the reason that it is considerably larger than any (unit-equivalent) collapse time observed at the parameter resolutions probed without being ungainly. To give an idea, the longest time to collapse in the scenarios graphed in Fig.~\ref{fig:combinedcolor} is $\approx 160$. It is by the apparent failure to collapse, the decline in field amplitude, and the recovery of the lapse to $\approx 1.0$ at such large times that a given configuration may be safely deduced to tend toward asymptotic flatness.
\par We use standard adaptive step size with Richardson extrapolation and multigrid techniques~\cite{obligatory} to greatly reduce computation time near criticality, without which a computationally prohibitive number of gridpoints would be required for accurate evolution.
\section{Results}
\subsection{Convergence}
\par
\begin{figure}[!ht]
\centering
\captionsetup{justification=centering}
\includegraphics[scale=0.32]{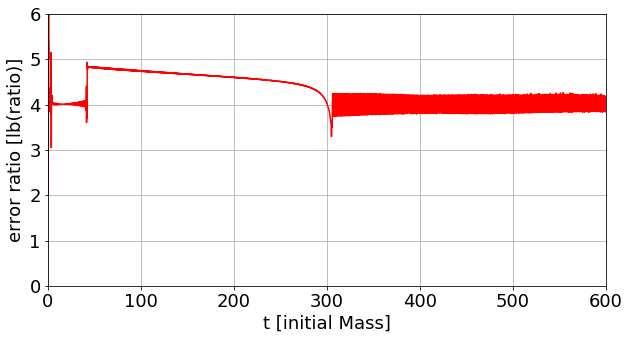}
\caption{\label{fig:shortdisperse}Approximate order of convergence given as a ratio of rms momentum constraint violation for 4000 and 8000 coarse gridpoints over time for disperse scenario. Initial conditions are somewhat close to criticality. Numerical boundary is at r=300. The jump in convergence order at $t~\approx~40$ is a consequence of a secondary grid activating.}
\end{figure}

 We first provide evidence of the expected fourth-order convergence, illustrated by the following figures. Taking a field configuration asymptotically dispersing with numerical boundary at $r=300$, we plot in Fig.~\ref{fig:shortdisperse} an example of the behavior of the apparent order of convergence as measured by relative momentum constraint violation using
\begin{equation}
{\rm error~ratio} = \ln{\left(\frac{{\rm rms}(M_{4000})}{{\rm rms}(M_{8000})}\right)/\ln(2)}.
\end{equation}
\par
\begin{figure}[!ht]
\centering
\captionsetup{justification=centering}
\includegraphics[scale=0.32]{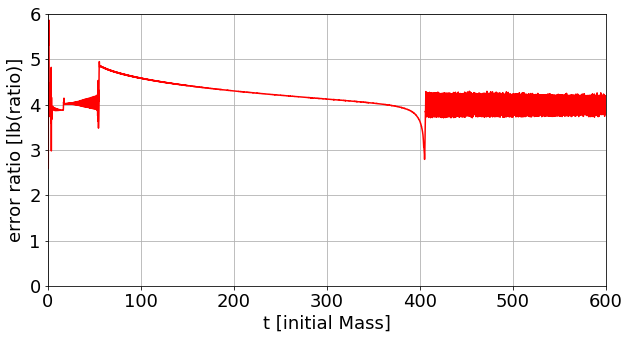}
\caption{\label{fig:longdisperse}Approximate order of convergence given as a ratio of rms momentum constraint violation for 4000 and 8000 coarse gridpoints over time for disperse scenario. Initial conditions are somewhat close to criticality. Numerical boundary is at r=400. The jump in convergence order at $t~\approx~60$ is a consequence of a secondary grid activating.}
\end{figure}

\begin{figure}[!ht]
\centering
\captionsetup{justification=centering}
\includegraphics[scale=0.32]{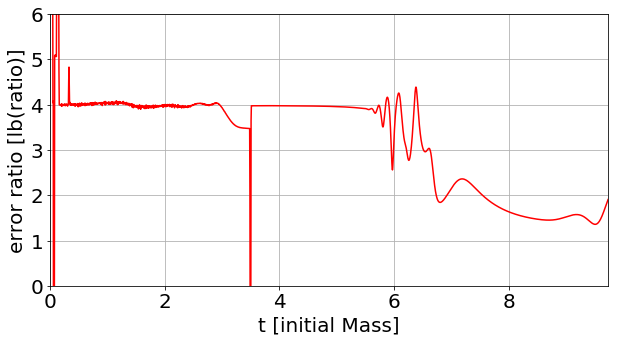}
\caption{\label{fig:shortcollapse}Approximate order of convergence given as a ratio of rms momentum constraint violation for 4000 and 8000 coarse gridpoints over time for collapse scenario. Numerical boundary is at $r~\approx~40$. The apparent dip at $t~\approx~3$ is due to a secondary grid appearing surrounding the origin. The black hole begins to form around $t~\approx~6$. Initial conditions are somewhat close to criticality.}
\end{figure}

 The behavior of the same setup, except with the numerical boundary stretched to $r=400$, is plotted in Fig.~\ref{fig:longdisperse}. In this second graph the apparent order of convergence does not exhibit significant oscillations until the elapsed asymptotic time matches the extended radial boundary -- this is typical behavior. Finally, the logarithmic error ratio for a configuration whose initial conditions are such that collapse occurs (with boundary at a much smaller $r=40$) is shown in Fig.~\ref{fig:shortcollapse}. The early decline in order in this last case may be attributed to two causes: the simultaneously numerically and physically significant fact that the polar areal gauge is not able to effectively evolve spacetimes for long after black hole formation, and the purely numerical fact that the differential equations for the metric become increasingly stiff as the lapse collapses.
\par In all graphs the ordinate value, being a measure of the apparent order observed, should be $\approx 4$ or $5$ (courtesy of Richardson extrapolation and multigridding), provided the scheme is stable and converging. The apparent satisfaction illustrated for most of the time evolution suggests that our algorithm exhibits convergence, and hence that the results obtained are not numerical artifacts.

\subsection{Scale interaction}
\par
\begin{figure}[!ht]
\centering
\begin{subfigure}{.23\textwidth}
\includegraphics[width=\linewidth]{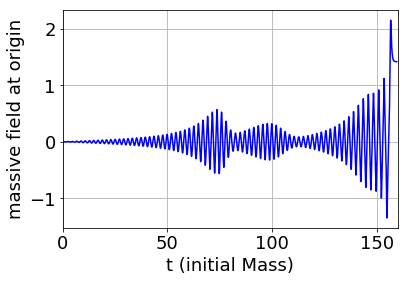}
\caption{}\label{subfig:f0011120a}
\end{subfigure}
\hspace*{\fill}
\begin{subfigure}{.23\textwidth}
\includegraphics[width=\linewidth]{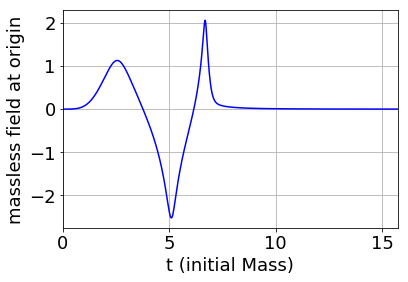}
\caption{}\label{subfig:f00434zoom}
\end{subfigure}\\
\begin{subfigure}{.23\textwidth}
\includegraphics[width=\linewidth]{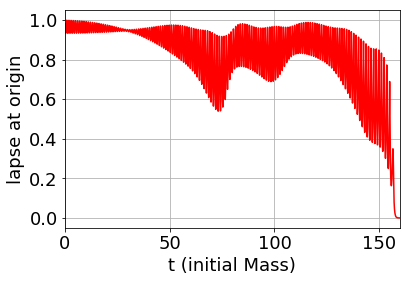}
\caption{}\label{subfig:al0011120a}
\end{subfigure}
\hspace*{\fill}
\begin{subfigure}{.23\textwidth}
\includegraphics[width=\linewidth]{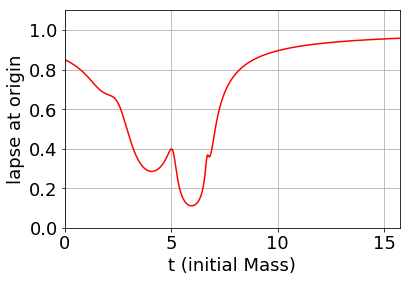}
\caption{}\label{subfig:al00434zoom}
\end{subfigure}
\caption{\label{fig:individualfields}Graphs of $\Phi$ (top) and $\alpha$ (bottom) at the origin for Type~I and type~II critical fields. The massive field, in the left column, is supercritical, while the massless field on the right is subcritical.}
\end{figure}

\begin{figure}[!ht]
\centering
\begin{subfigure}{0.4\textwidth}
\includegraphics[width=\linewidth]{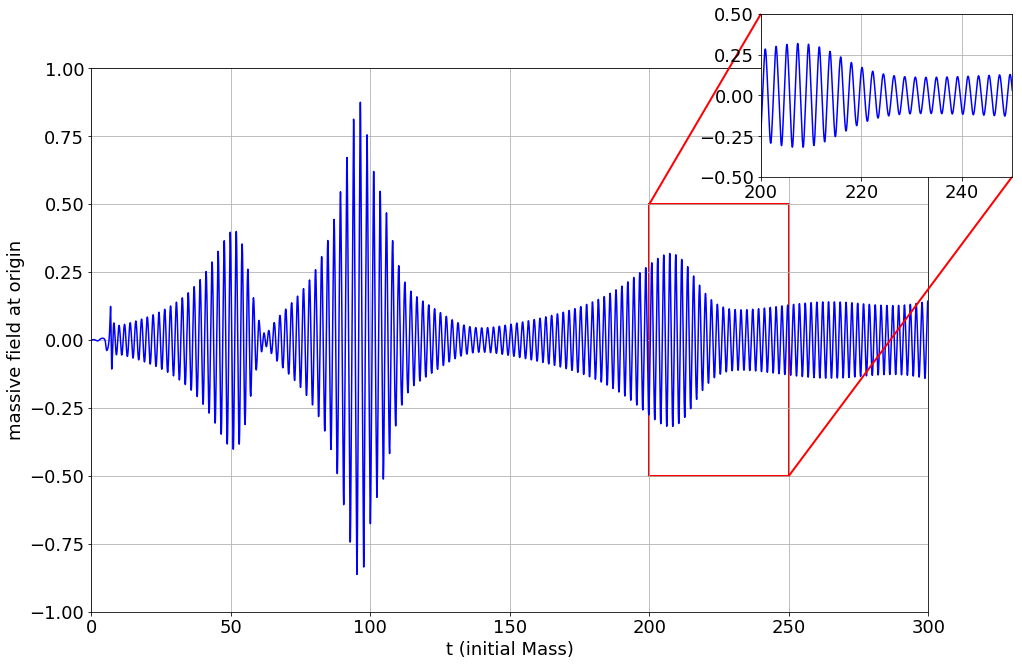}
\caption{}\label{subfig:f20011120424}
\end{subfigure}\\
\begin{subfigure}{0.4\textwidth}
\includegraphics[width=\linewidth]{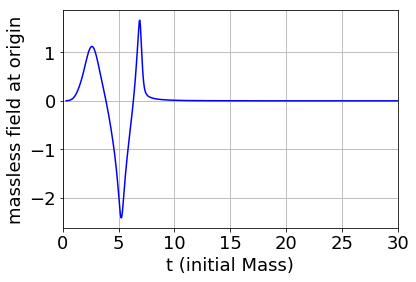}
\caption{}\label{subfig:f0011120434}
\end{subfigure}\\
\begin{subfigure}{0.4\textwidth}
\includegraphics[width=\linewidth]{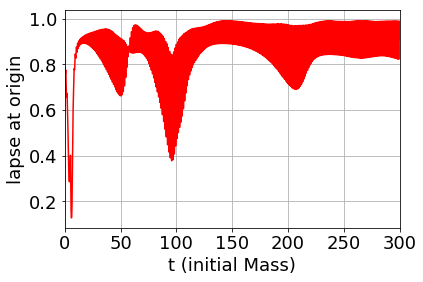}
\caption{}\label{subfig:al00011120434}
\end{subfigure}
\caption{\label{fig:mixedfields}Graphs of the $\Phi$s and $\alpha$ at the origin for mixed field content. The inset for (a), depicting the massive field, illustrates how rapidly the field varies. No collapse occurs, despite the initial field content being a combination of the two fields whose time evolutions are depicted individually in Fig.~\ref{fig:individualfields} \textit{supra}.}
\end{figure}

 The presence of two near-critical fields associated with different critical spacetimes results in competition between their respective evolutionary tendencies. We first consider individually, as a concrete example, a massless field configured to disperse and a massive field configured to collapse, both with Gaussian-like initial data. We plot the resulting behaviors of the fields and the lapse at the origin in Fig.~\ref{fig:individualfields}. Figure~\ref{fig:mixedfields}, meanwhile, depicts the behavior of the two fields when they are simultaneously present, coupled only by their mutual minimal coupling to gravity. The apparent space-filling seen in many of these graphs, both for the fields and the lapse, arises not from numerical error, but rather is the result of rapid oscillations: this is shown by the inset in the first image of Fig.~\ref{fig:mixedfields}. This is an expected consequence of the extra timescale introduced by the intrinsic field mass -- the massless case notably does not feature such rapid variation.
\par It is notable is that \textit{no} collapse occurs in Fig.~\ref{fig:mixedfields}, despite the initial data having greater mass-energy content than either near-critical constituting field taken alone. This surprising result suggests that the two fields frustrate, rather than enhance, their respective critical evolutions. In this strongly-coupled system we are seeing nonlinear phenomena overruling common intuition.

\par
\begin{figure}[!ht]
\centering
\begin{subfigure}{0.23\textwidth}
\includegraphics[width=\linewidth]{f0011120.png}
\caption{}\label{subfig:f0011120b}
\end{subfigure}
\hspace*{\fill}
\begin{subfigure}{0.23\textwidth}
\includegraphics[width=\linewidth]{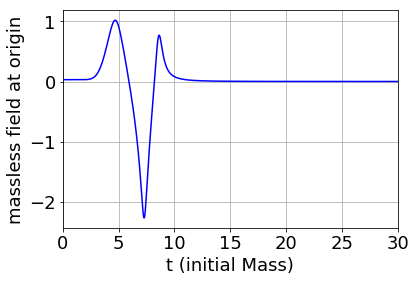}
\caption{} \label{subfig:f034zoom}
\end{subfigure}\\
\begin{subfigure}{0.23\textwidth}
\includegraphics[width=\linewidth]{al0011120.png}
\caption{} \label{subfig:al0011120b}
\end{subfigure}
\hspace*{\fill}
\begin{subfigure}{0.23\textwidth}
\includegraphics[width=\linewidth]{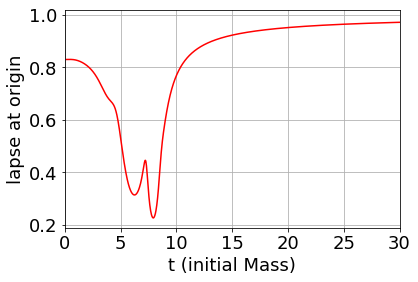}
\caption{} \label{subfig:al034zoom}
\end{subfigure}
\caption{\label{fig:individualfields+}$\Phi$ (top) and $\alpha$ (bottom) at the origin with hyperbolic tangent initial data for the massless field. The massive field, in the left column, is supercritical, while the massless field on the right is subcritical. }
\end{figure}

\begin{figure}[!ht]
\centering
\begin{subfigure}{0.4\textwidth}
\includegraphics[width=\linewidth]{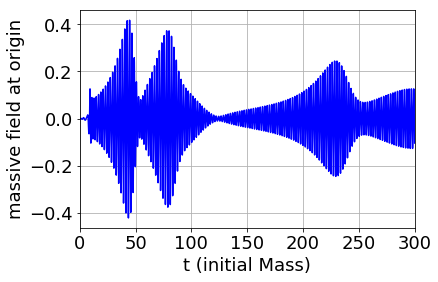}
\caption{} \label{subfig:f2001112034}
\end{subfigure}\\
\begin{subfigure}{0.4\textwidth}
\includegraphics[width=\linewidth]{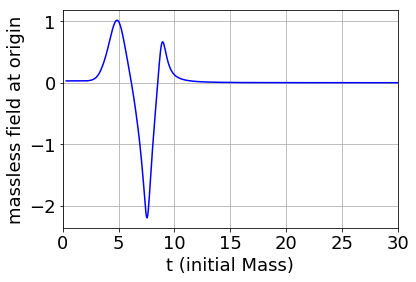}
\caption{} \label{subfig:f0001112034}
\end{subfigure}\\
\begin{subfigure}{0.4\textwidth}
\includegraphics[width=\linewidth]{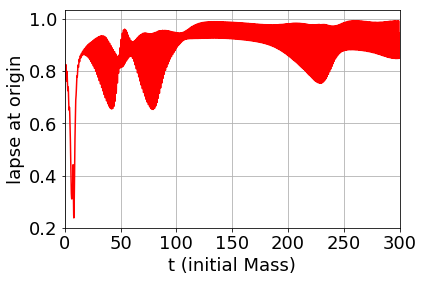}
\caption{} \label{subfig:al0001112034}
\end{subfigure}
\caption{\label{fig:mixedfields+}Graphs of $\Phi$ and $\alpha$ at the origin with hyperbolic tangent initial data for the massless field. No collapse occurs despite the initial field content being a combination of the two fields whose time evolutions are depicted individually in Fig.~\ref{fig:individualfields+} \textit{supra}.}
\end{figure}

 Nor is this oddity dependent upon any quirk of the initial data, as should be expected given the underlying quasi-universality. Taking the massless field to be a shifted hyperbolic tangent function as its initial data yields similar results. Figure~\ref{fig:individualfields+} illustrates the behavior of each field alone, while Fig.~\ref{fig:mixedfields+} shows the evolution of the two taken at once.

\par
\begin{figure*}[!htp]
\centering
\includegraphics[scale=0.45]{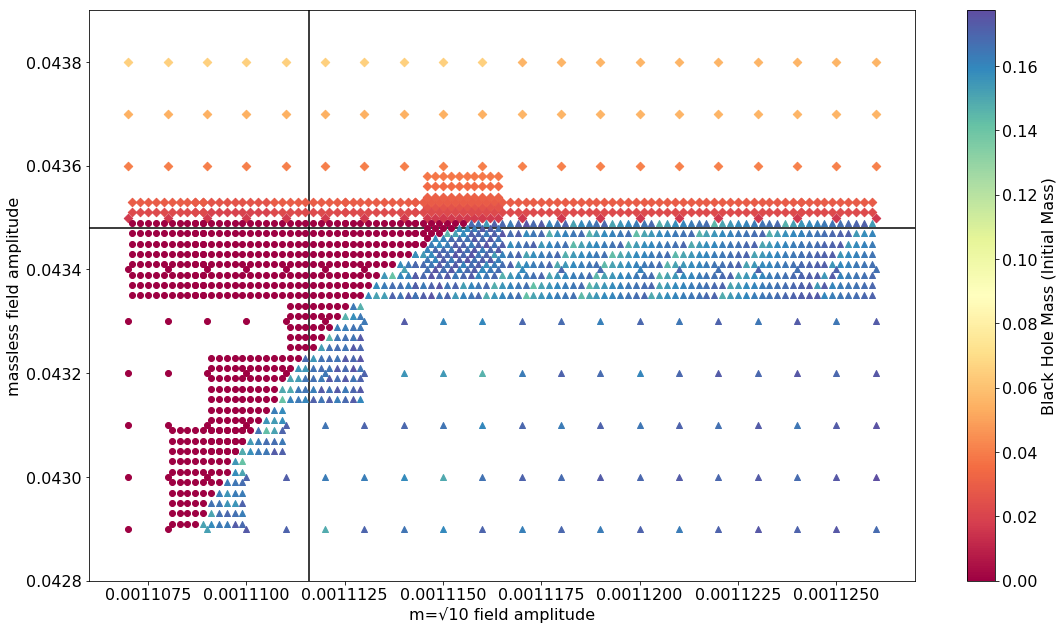}
\caption{\label{fig:combinedcolor} Phase diagram of time evolution behavior for the multicritical field configuration considered in this paper. The axis variables correspond to the parameters tuning the initial data for the scalar fields that ultimately determine whether collapse or dispersal occurs. The circles indicate dispersal, the triangles denote type~I collapse, and the diamonds are type~II -- this is determined by collapse times (or the lack thereof). The scale is a measure of the mass of the black hole formed. If the initial data evolves to be asymptotically dispersing, this is set to zero mass. The approximate critical quantities for the individual fields are$~0.04347$ and$~0.0011116$. The above picture suggests that the competition between the two associated critical spacetimes considered here has an inhibiting effect on criticality.}
\end{figure*}

 On a more comprehensive level, Fig.~\ref{fig:combinedcolor} depicts a kind of phase diagram we have obtained for the asymptotic behaviors of our composite mulitcritical configurations. Each individual point represents an independent simulation, with the abscissa and ordinate values specifying the amplitudes for the massive and massless initial fields respectively. The marker shapes classify the spacetimes by apparent end behavior, distinguishing dispersal (circles), type~I collapse (triangles), and type~II collapse (diamonds). The method utilized for this classification is crude, but sufficient: collapses occurring $<40$ time units are classified as type~II, collapses occurring thereafter up to $t=400$ time units are classified as type~I, and spacetimes showing no signs of collapse up to $t=400$ are deemed asymptotically dispersing. This cutoff time is well more than necessary, since the greatest collapse times occur at $\approx 160$ time units at the parameter resolution probed. Meanwhile, the scale applied to the points reflects the black hole mass at the time of collapse, set to zero for dispersing spacetimes. Three distinct domains emerge in both classification schemes, which are found to be in complete agreement with each other. 
\par The solid black horizontal and vertical lines in the same figure denote the approximate critical parameter for two fields if they were taken alone. Figure~\ref{fig:combinedcolor} shows, however, that the three domains are not circumscribed by these lines, as one might assume. The asymptotically dispersing domain is raised slightly into the would-be type~II critical region, and also bent rather noticeably into what might naively be taken to be the type~I supercritical region. Our specific scenario hence shows that multi critical configurations can actually have a consistently \textit{inhibiting} influence on black hole formation.

\section{Discussion}
\par In a sense, there are really only two domains in Fig.~\ref{fig:combinedcolor} if the configurations are considered in the asymptotic time limit: either a single black hole forms, or the fields disperse and spacetime tends toward flatness asymptotically. The spacetimes exhibiting early collapse feature black holes that grow as time progresses, courtesy of the second still-ingoing field. This would be seen if more delicate evolution techniques were employed, albeit at far greater computational cost. Nevertheless, the vastly differing collapse timescales seen, in conjunction with the encroachment of the asymptotically dispersing region, suggest that different dynamics -- that is, different relevant modes -- are responsible for steering time evolution within these three domains. The sharp boundaries between the apparent regions seen in Fig.~\ref{fig:combinedcolor} are manifestations of the structure of the relevant system of attractors at play in our scenario.
\par
\begin{figure}[!ht]
\centering
\includegraphics[scale=0.38]{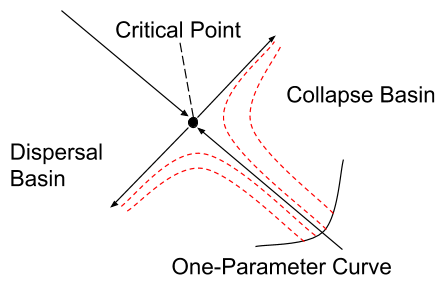}
\caption{\label{fig:dynosimple}Simplified picture of the dynamics of critical phenomena for a single field. The dashed trajectories denote the time evolution of spacetimes with initial conditions on the parametrized curve. The arrows suggest the direction of the locally dominant time evolution mode: a single attracting surface, here represented as a point, is attractive (has arrows pointing to it) on a submanifold of some codimension. In directions normal to this submanifold it is repulsive (has arrows pointing away from it), so a one-parameter line of initial data not lying exactly in this submanifold may have points close to intersection with vastly differing asymptotic behaviors.}
\end{figure}

 We suggest a simple dynamical systems picture for understanding this effect. As is well known~\cite{gundreview} for a single field -- and simply sketched in Fig.~\ref{fig:dynosimple} -- critical collapse in general relativity is the consequence of the existence of an attractor of some codimension in the phase space of solutions to Einstein's equations. A generic one-parameter curve of initial data intersects the surface of attraction at a single point corresponding to the critical value. Initial data given by configurations described by a parameter slightly greater or less than criticality will, after possibly lengthy critical evolution, be repelled in opposite directions from the surface of attraction toward different asymptotic limits -- either black hole formation, or dispersal to flat space. 

\par When two or more near-critical fields are in play at once, however, and the two fields are configured to have their evolutions determined by two different critical surfaces, then a more complex picture could emerge in which the various attracting surfaces are in a sense in competition. As a consequence, the stronger attractors (heuristically corresponding to the critical surface with larger inverse timescale, corresponding to the smaller mass solution -- the massless type~II critical solution in our case) will ``pull'' initial data away from other attractors, possibly resulting in dispersal to flat space for some configurations despite being supercritical with respect to at least one of the parameters. This effect is observed in our phase space picture Fig.~\ref{fig:combinedcolor} cohabitant with significant curvature of the domain separation, which supports this interpretation. What is surprising here is the dispersal of spacetimes where a black hole would form but for the presence of a competing field. 

\begin{figure}[!ht]
\centering
\includegraphics[scale=0.38]{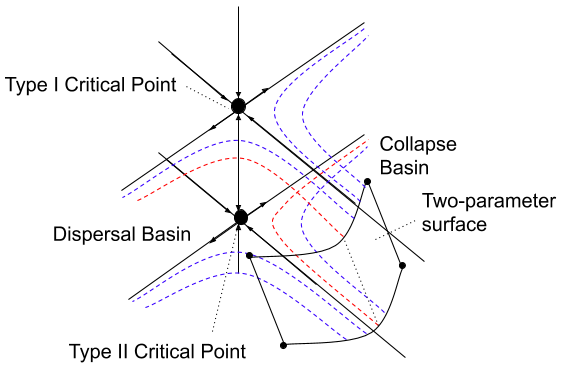}
\caption{\label{fig:dynobad}Simplified picture of the possible dynamics for competitive critical phenomena. The dashed trajectories denote the time evolution of spacetimes with initial conditions on the parametrized surface, while the arrows suggest the direction of locally dominant time evolution modes. Because the influence of two different attractors is relevant, there is a kind of competition between their effects as initial data is ``pulled'' toward both surfaces, in a sense inhibiting the criticality of both. The pair of dashed trajectories denote the time evolution of initial fields having equal massless field parameters, but different massive field parameters. This shows how attraction to a secondary critical point may nontrivially inhibit a configuration's tendency to collapse.}
\end{figure}

\par Our results should be compared with a recent, earlier paper by Gundlach, Baumgarte, and Hilditch~\cite{panic}. In their paper, they consider the interaction of an SU(2) Yang-Mills field with a massless scalar field, with the intent of investigating what effects gravitational waves might have on critical collapse. They find that the scalar field (acting as a toy model of gravitational waves) in fact dominates on smaller scales, and postulate the existence of a family of ``quasi-discretely self-similar'' spacetimes with one unstable mode that controls the evolution of their mixed field configuration. These postulated solutions interpolate between the critical spacetimes of the individual field constituents, moving from pure Yang-Mills in the distant past to pure scalar in the distant future. 
\par Gundlach \textit{et al.} explain this scenario from a dynamical systems perspective, with one of the Yang-Mills critical solution's unstable modes directed toward the critical scalar solution, which has but a single unstable mode. This picture, illustrated in Fig. 12 of their paper, loosely resembles the scenario we conjecture in Fig.~\ref{fig:dynobad} of ours. However, they also briefly suggest the existence of an alternative scenario for other mixed field configurations (they give the example of two massless scalar fields) with \textit{three} critical spacetimes: one for each of the two constituent fields, each with a single unstable mode, and a third with a pair directed toward the other critical solutions. Our case more closely exhibits this latter scenario, with the third hypothetical critical solution positioned along the frustrated axis containing the type~I and type~II critical points in Fig.~\ref{fig:dynobad}. If such a spacetime exists, then our conjectured scenario very much resembles Fig. 13 of their paper. Precisely this is suggested by Fig.~\ref{fig:combinedcolor}. Each boundary is indicative of an unstable mode directed away from a particular attractor of some codimension. There appears to exist a kind of triple point, deviation from which along two of the boundaries leads to what would appear to be the type~I and type~II critical solutions. Movement along the third boundary between the type~I and type~II regions, meanwhile, is in fact only movement toward generic asymptotic black hole solutions, as explained at the beginning of this section. This last boundary is hence symptomatic not of another family of critical solutions in addition to the triple point, but rather the different modes dominating time evolution. 
\par It is interesting, nevertheless, how different the time scales of the ``critical'' evolution are on either side. We interpret this to be a consequence of the vastly differing criticality types investigated. It is possible that the alternate scenario alluded to in Gundlach \textit{et al.}'s paper containing two massless scalar fields might show a similarly exaggerated difference in timescale if, for example, both fields were taken to be initially thin shells, with one field localized at a significantly greater radius. Such a configuration, however, would likely not exhibit the same mass behavior at the time of collapse which so readily illustrates which modes dominate time evolution for a given initial datum. Moreover, it seems likely that a configuration with two massless scalar fields, or more generally two fields associated to the same type of criticality, would rather enhance criticality, decreasing the critical value along either axis of the two parameter space (assuming both parameters to be positively correlated with energy density). This, if true, would contrast with our scenario, which exhibits instead the inhibiting influence of its form of multi-criticality.
\section{Conclusion}
We have found that the evolution of initial data containing multiple fields tuned near-criticality with respect to distinct critical surfaces exhibits a kind of competition between the critical surfaces. On a higher level, this is consistent with the results of a recent paper by Gundlach \textit{et al.}~\cite{panic}, though we employ different methods and analyze a different scenario. This behavior is expected of the Einstein equations if they are approached with the philosophy of dynamical systems. Though this paper only made use of two scalar fields, it is likely that this phenomenon generalizes for the case of more fields and more varied matter content. Moreover, we have no reason not to expect other dynamical systems phenomena, such as bifurcation, to manifest in other regions of the parameter space of initial conditions away from multipoints. Outside the well-behaved region containing the triple point seen in Fig.~\ref{fig:combinedcolor}, we have found more complex behavior near the boundary between type~I collapse and dispersal. This is the subject of ongoing study.
\par This interaction may have implications for cosmology in the production of PBHs. The effect observed would seem to indicate that combined matter configurations may in fact at times inhibit critical formation, which necessitates a more delicate treatment of fluctuations when applying critical black hole phenomena to PBHs.
\begin{acknowledgments}
We thank Philip Beltracci, Benjamin Bromley, Eric Hirschmann and David Neilsen for valuable discussions, and Tugdual LeBohec for the donation of processor time used in the calculations presented here. We also gratefully acknowledge the technical support of the Center for High Performance Computing at the University of Utah.
\end{acknowledgments}

\bibliography{ms}
\end{document}